\documentstyle[12pt]{article}
\newlength{\extraspace}
\newlength{\extraspaces}
\newcommand{\be}{\begin{equation}
\addtolength{\abovedisplayskip}{\extraspaces}
\addtolength{\belowdisplayskip}{\extraspaces}
\addtolength{\abovedisplayshortskip}{\extraspace}
\addtolength{\belowdisplayshortskip}{\extraspace}}
\newcommand{\ee}{\end{equation}}
\newcommand{\ba}{\begin{eqnarray}
\addtolength{\abovedisplayskip}{\extraspaces}
\addtolength{\belowdisplayskip}{\extraspaces}
\addtolength{\abovedisplayshortskip}{\extraspace}
\addtolength{\belowdisplayshortskip}{\extraspace}}
\newcommand{\ea}{\end{eqnarray}}
\newcommand{\nonu}{\nonumber \\[.5mm]}
\newcommand{\A}{&\!\!\!}
\begin{document}
\thispagestyle{empty}
\setlength{\baselineskip}{6mm}
\begin{flushright}
SIT-LP-01/13\\
January, 2013
\end{flushright}
\vspace{7mm}
\begin{center}
{\large \bf Chiral Symmetry}
\\[15mm]
{\sc Kazunari Shima}
\footnote{
\tt e-mail: shima@sit.ac.jp} \ 
and \ 
{\sc Motomu Tsuda}
\footnote{
\tt e-mail: tsuda@sit.ac.jp} 
\\[2mm]
{\it Laboratory of Physics, 
Saitama Institute of Technology \\
Fukaya, Saitama 369-0293, Japan} \\[7mm]
{\sc Takeshi Okano}
\footnote{
\tt e-mail: okano@sit.ac.jp} 
\\[2mm]
{\it Department of Mathematics, 
Saitama Institute of Technology \\
Fukaya, Saitama 369-0293, Japan} \\[15mm]

\begin{abstract}
The exact classical solution of the equation of the motion for the Nambu-Goldstone fermion of the nonlinear 
representation of supersymmetry and its physical significance are discussed, 
which gives a new insight into the chiral symmetry of the standard model for the low energy particle physics. 
\\[7mm]
%
%
\end{abstract}
\end{center}

\newpage

\noindent
Supersymmetry (SUSY) \cite{WB}, which constitutes the space-time symmetry treating 
the fermionic and the bosonic degrees of freedom on an equal footing  
is recognized as the key notion for the unified description of space-time and matter. 

The linear representation of SUSY (LSUSY) \cite{WZ1} is realized on the multiplet of fields. 
Various models based upon $N=1$ LSUSY have been studied extensively 
and remarkable phenomenological and field theoretical results have been obtained \cite{WB}.    
The simple version of the minimal supersymmetric standard model (MSSM) 
with the accessible low TeV  SUSY breaking, which may stabilize the 
mass of light Higgs particle, looks severely constrained by the LHC experiment. 
The unpleasant point of LSUSY model is that the mechanism and the physical meaning 
of the spontaneous SUSY breaking, e.g., the origin of $D$ term, is not clear.

While, the nonlinear representation of SUSY (NLSUSY) \cite{VA} is written down by the Nambu-Goldstone 
fermion field and realized {\it geometrically} 
on specific ${\it flat}$ space-time equipped with the robust spontaneous SUSY breaking encoded 
in the nature of space-time itself. 
The action containing higher orders of the stress-energy-momentum tensors 
of the fermion field is almost unique up to the higher-order derivative terms. 
Despite the apparent non-renormalizability it may be worth exploring in the NLSUSY framework 
a new paradigm for the unified theory of nature in terms of the fermion field, 
which is the long standing challenge.

To incorporate the space-time dynamics we assume that the ultimate shape of nature is 
the {\it four} dimensional (Riemann) manifold 
possessing the NLSUSY degrees of freedom on {\it tangent} space, i.e., 
{\it tangent} space is specified by the Grassmann coordinates $\psi_{\alpha}$ 
besides the Minkowski coordinates $x_a$. 
NLSUSY can be easily fused with the general relativity (GR) principle, which 
produces the nonlinear supersymmetric general relativity theory (NLSUSYGR) \cite{KST,QTS-hepth-procs} 
in the form of Einstein-Hilbert (EH) action of the ordinary GR 
equipped with the NLSUSY cosmological term indicating the NLSUSY nature of tangent space.  
NLSUSYGR {\it would} break down spontaneously ({\em Big Decay}) to EH action 
for ordinary Riemann space-time (graviton) and NLSUSY action 
for massless Nambu-Goldstone (NG) fermion ({\it superon}) called {\it superon graviton model (SGM)} corresponding to 
the spontaneous space-time SUSY breaking; [super $GL(4,{\bf R})$/$GL(4,{\bf R})$] \cite{KST,QTS-hepth-procs}. 
Although NLSUSY is the non-renormalizable highly nonlinear fermionic theory,  
it is shown by the systematic linearizations of NLSUSYGR in flat space-time that 
the NLSUSY cosmological term is recasted (equivalent) to the familiar broken 
LSUSY theory, i.e., the familiar broken LSUSY theory emerges in the true vacuum of NLSUSY theory 
as the low energy (effective) theory, 
where all fields including $D$ term are composed of NG fermion degrees of freedom \cite{QTS-hepth-procs,STL}.
NLSUSYGR scenario has a potential to give new insights into the SUSY effects 
which can be tested  in cosmology and 
the low energy particle physics \cite{QTS-hepth-procs,STL}. 

In this short letter we study the exact classical solution of the equation of the motion for NLSUSY, 
which may explain the chirality of all fermions in the SM (MSSM) for the low energy particle physics. 

The $N = 1$ NLSUSY action $L_{\rm NLSUSY}$ in flat space-time 
is given by Volkov and Akulov as follows \cite{VA}; 
\begin{equation}
L_{\rm NLSUSY} = -{1 \over {2 \kappa^2}} \vert w \vert 
=  -{1 \over {2 \kappa^2}} \left[ 1 + t^a{}_a 
+ {1 \over 2} (t^a{}_a t^b{}_b - t^a{}_b t^b{}_a) 
+ \cdots \right], 
\label{VA}
\end{equation}
where 
\begin{equation}
\vert w \vert = \det w^a{}_b = \det(\delta^a_b + t^a{}_b), \ \ 
t^a{}_b =- i \kappa^2 \bar\psi \gamma^a \partial_b \psi 
\label{detw}
\end{equation} 
and $\psi$ is a four components Majorana spinor, 
which balances the fermionic and the bosonic degrees of freedom. 
\footnote{
Minkowski space-time indices are denoted by $a, b, \cdots = 0, 1, 2, 3$ 
and the flat metric is $\eta^{ab} = {\rm diag}(+,-,-,-)$. 
Gamma matrices satisfy $\{ \gamma^a, \gamma^b \} = 2 \eta^{ab}$ 
and we define $\sigma^{ab} = {i \over 4}[\gamma^a, \gamma^b]$. 
}
$L_{NLSUSY}$ is invariant under $N=1$ NLSUSY transformation, 
\begin{equation}
\delta_\zeta \psi = {1 \over \kappa} \zeta - i \kappa \bar\zeta \gamma^a \psi \partial_a \psi,  
\label{nlsusy}
\end{equation}
where $\zeta$ is a global Majorana spinor parameter. 
$\psi$ is the coordinates of the coset space ${Super-Poincare \over  Poincare}$ 
and subsequently recasted as the NG fermion in the vacuum of $L_{\rm NLSUSY}$. 
$\kappa$ with the dimension $[length]^{+2}$ gives 
the strength of the coupling of NG fermion to the vacuum. 

The variation of (\ref{VA}) with respect to $\bar\psi$ gives the following 
equation of motion for $\psi$; 
\ba
\A \A 
\!\!\not\!\partial \psi 
- i \kappa^2 \left\{ T^a{}_a \!\!\not\!\partial \psi - T^a{}_b \gamma^b \partial_a \psi 
+ {1 \over 2} (\partial_a T^b{}_b - \partial_b T^b{}_a) \gamma^a \psi \right\} 
\nonu
\A \A 
- {1 \over 2}(-i \kappa^2)^2 \epsilon_{abcd} \epsilon^{efgd} 
(T^a{}_e T^b{}_f \gamma^c \partial_g \psi + T^a{}_e \partial_g T^b{}_f \gamma^c \psi) 
= 0,   
\label{Eqn}
\ea
where $T^a{}_b = i \kappa^{-2} t^a{}_b = \bar\psi \gamma^a \partial_b \psi$. 
The highest order term with ${\cal O}(T^4)$ in the NLSUSY action (\ref{VA}) 
is known to vanish identically for $N = 1$ SUSY \cite{KM}. 

It is not easy to solve (\ref{Eqn}) in general. 
However, considering $\psi$ is the NG fermion it is natural to argue the (free) massless solution, 
namely, we consider the case, 
\be
\!\!\not\!\partial \psi(x) = 0, 
\label{npla}
\ee
which eliminates the lowest order term of (\ref{Eqn}). 

In NLSUSYGR scenario, the NG fermion $\psi(x)$ is besides $x^a$ 
the Grassmann degrees of freedom on tangent space of ultimate space-time $(x^a, \psi \ ; x^\mu)$. 
This means that the exact solution of $\psi(x)$ is dictated by the nontrivial geometry of space-time, 
even in asymptotic flat space-time. 
In fact, we have shown in supergravity \cite{FNF,DZ} that the nontrivial exact classical solution of 
the massless (free) gravitino under the Coulomb type condition $\psi_\mu = \delta_{0 \mu} \psi$ 
exists in flat space-time of the Kerr solution of GR, not in that of Schwarzshild solution \cite{SK}. 
Therefore, the condition (\ref{npla}) is worth considering seriously from the physical point of view, 
which allow the non-trivial (localized) solution depending on the geometry of space-time. 

We will see throughout the discussion that Eq.(\ref{Eqn}) is satisfied identically at each order in $\kappa$, 
provided $\psi$ is {\em a single} chiral state which satisfies (\ref{npla}), i.e. \ 
$\!\!\not\!\partial \psi_{L}(x) = 0$ {\em or} $\!\!\not\!\partial \psi_{R}(x) = 0$. 
We show below some details of the computation. 

First let us study the term, $T^a{}_b \gamma^b \partial_a \psi$, 
in the second term at ${\cal O}(\kappa^2)$ of Eq.(\ref{Eqn}), 
for the first and the third terms at ${\cal O}(\kappa^2)$ 
are explicitly proportional to \ $\!\!\not\!\partial \psi$. 
By using the Fierz transformation and considering the (anti)commuting properties 
for the Majorana spinor and Eq.(\ref{npla}), 
many terms including belowmentioned mass-like terms vanish manifestly. 
Consequently, the following (axial)vector-type terms survive in the second term;
%
\be
T^a{}_b \gamma^b \partial_a \psi \sim 
{i \over 2} \epsilon^{abcd} (\partial_a \bar\psi \gamma_b \psi \ \gamma_5 \gamma_c \partial_d \psi 
- \partial_a \bar\psi \gamma_5 \gamma_b \psi \ \gamma_c \partial_d \psi). 
\label{example}
\ee
Again by the Fierz transformation the terms (\ref{example}) become mass-like terms, 
\be
-{i \over 4} \epsilon^{abcd} (\partial_a \bar\psi \gamma_b \gamma_e \partial_d \psi \ \gamma_5 \gamma_c \gamma^e \psi 
- \partial_a \bar\psi \gamma_5 \gamma_b \gamma_e \partial_d \psi \ \gamma_c \gamma^e \psi), 
\ee
which vanish for the single chiral state. 

Next we discuss the terms at ${\cal O}(\kappa^4)$ in Eq.(\ref{Eqn}). 
From the relation, 
\ba
\A \A 
\epsilon_{abcd} \epsilon^{efgd} T^a{}_e T^b{}_f \gamma^c \partial_g \psi 
\nonu
\A \A 
= -(T^a{}_a T^b{}_b - T^a{}_b T^b{}_a) \!\!\not\!\partial \psi 
- 2(T^a{}_c T^c{}_b - T^c{}_c T^a{}_b) \gamma^b \partial_a \psi, 
\label{example2}
\ea
we give an example of calculations for $T^a{}_c T^c{}_b \gamma^b \partial_a \psi$ 
in the third term of Eq.(\ref{example2}). 
In this case the following term survives by using repeatedly the Fierz transformation, 
\be
T^a{}_c T^c{}_b \gamma^b \partial_a \psi \sim 
{1 \over 4} \gamma_a \sigma_{de} \sigma_{fg} \partial^a \psi 
\ \bar\psi \gamma_b \sigma^{de} \partial^b \psi \ \bar\psi \gamma_c \sigma^{fg} \partial^c \psi. 
\label{example3}
\ee
However, it can be shown that the term (\ref{example3}) is expressed 
by means of only terms proportional to \ $\!\!\not\!\partial \psi$ 
by using Clifford algebra of $\gamma$ matrices. 
As for the last term in Eq.(\ref{Eqn}), the similar argument shows that 
it vanishes identically for the single chiral state satisfying Eq.(\ref{npla}). 

This means that the NG fermion with higher order self-interactions of NLSUSY is the massless chiral fermion 
which satisfies either of the equation $\!\!\not\!\partial \psi_L(x) = 0$ or $\!\!\not\!\partial \psi_R(x) = 0$. 
This situation is interesting, for the lowest order of $T_{a}{}^{b}$ , 
i.e., the kinetic term does not constrain the chirality of the (left-right symmetric) 
solution by itself, however higher order self-interaction terms with 
NLSUSY structure $as$ $a$ $whole$  constrain the chirality of the solution 
to $one$ (\,left\, $or$\, right\,) of two chiral eigenstates. 

If NLSUSYGR is the underlying principle beyond the SM, this mechanism  
may explain the chiral eigenstate for all massless fermions in the SM, 
which is a basic assumption within the SM framework. 
Considering that the observed fermion in the SM is inconsistent with the NG fermion, 
the SM may be the composite (effective) theory of the NG fermion as in NLSUSYGR scenario. 
In fact, in $N=2$ minimal SGM scenario ($N=1$ is unphysical \cite{STT}),   
LSUSY QED (and probably the MSSM as well) emerges in the $true$ vacuum of NLSUSY, 
where all particles for the LSUSY multiplet are composites of superons. 
For example, the massless fermion $\lambda^I$ ($I=1,2$ for leptons and quarks) in Wess-Zumino gauge
is composed of superon $\psi^I$ as follows \cite{STL}; 
\begin{equation}
\lambda^I \,\,\!\!\! =  \left\{ \psi^I \vert w \vert 
- {i \over 2} \kappa^2 \partial_a 
( \gamma^a \psi^I \bar\psi^J \psi^J \vert w \vert ) \right\}, 
\label{susyinv}
\end{equation} 
which shows that  quarks and leptons $\lambda^I$ are  ${\it composites}$ 
and ${\it chiral}$ fermions, because the constituent $\psi^I$ (superon) 
is chiral. 

In the SGM scenario of NLSUSYGR, the chiral symmetry (and SUSY as well) is broken 
spontaneously by 
the vacuum energy density of space-time (NLSUSYGR) recasted into the composite $D$ term. 
The ordinary  symmetry breaking for the SM (MSSM) occurs among corresponding composite fields. 
The SM (MSSM) may be regarded as the low energy effective theory of NLSUSYGR. 

Finally before conclusions, we  mention that NLSUSYGR (SGM) scenario constrains 
the dimensions of space-time to $four$, provided we require $SO(D-1,1) \cong SL(d,C)$, 
i.e., ${D(D-1) \over 2}=2(d^{2}-1)$  for the Lorentz invariance, 
which holds for only  $D=4,\,d=2$ besides the exceptional case $D=2, d=1$.   

Now we summarize the result as follows. 
The vacuum of NLSUSY predicts the massless chiral NG fermion (superon), 
where higher order self-interaction terms of the massless superon 
constrain the chirality of itself. 
The chiral eigenstate for quarks and leptons in the SM (MSSM), 
i.e. the chiral symmetry is understood naturally, 
if quarks and leptons are composite eigenstates of the chiral superon,  
which achieves simultaneously the true vacuum of NLSUSY as demonstrated 
in the SGM scenario of NLSUSYGR \cite{KST,QTS-hepth-procs}.

\newpage

%
\newcommand{\NP}[1]{{\it Nucl.\ Phys.\ }{\bf #1}}
\newcommand{\PL}[1]{{\it Phys.\ Lett.\ }{\bf #1}}
\newcommand{\CMP}[1]{{\it Commun.\ Math.\ Phys.\ }{\bf #1}}
\newcommand{\MPL}[1]{{\it Mod.\ Phys.\ Lett.\ }{\bf #1}}
\newcommand{\IJMP}[1]{{\it Int.\ J. Mod.\ Phys.\ }{\bf #1}}
\newcommand{\PR}[1]{{\it Phys.\ Rev.\ }{\bf #1}}
\newcommand{\PRL}[1]{{\it Phys.\ Rev.\ Lett.\ }{\bf #1}}
\newcommand{\PTP}[1]{{\it Prog.\ Theor.\ Phys.\ }{\bf #1}}
\newcommand{\PTPS}[1]{{\it Prog.\ Theor.\ Phys.\ Suppl.\ }{\bf #1}}
\newcommand{\AP}[1]{{\it Ann.\ Phys.\ }{\bf #1}}

\end{document}